\documentclass[5p, times, twocolumn]{elsarticle}

\usepackage{graphicx}

\journal{Computational Materials Science}
\begin{document}
\begin{frontmatter}
\title{Monte Carlo Simulations of Interacting Anyon Chains}

\author{Huan Tran and N. E. Bonesteel}
\address{Department of Physics and National High Magnetic Field Laboratory, Florida State University, Tallahassee, FL 32310, USA}

\begin{abstract}
A generalized version of the valence-bond Monte Carlo method is used to study ground state properties of the 1+1 dimensional quantum $Q$-state Potts models. For appropriate values of $Q$ these models can be used to describe interacting chains of non-Abelian anyons --- quasiparticle excitations of certain exotic fractional quantum Hall states.
\end{abstract}


\end{frontmatter}

\section{Introduction}

The notion of a valence bond, a simple singlet state formed by two localized spin-1/2 particles, captures much of the physics of how pairs of electrons are correlated in the real world.  This is certainly true when describing chemical bonds in molecules, where a local picture of electron correlations is clearly appropriate; however, the language of valence bonds has also proven useful for describing a variety of possible singlet ground states of quantum spin systems in the thermodynamic limit, including resonating valence-bond states and valence-bond solids.

Valence bonds can also be used for Monte Carlo simulations, as first shown in the context of variational Monte Carlo by Liang, Dou{\c c}ot and Anderson in 1988 \cite{lia88}.  More recently, Sandvik \cite{san05} has introduced a projector Monte Carlo method known as valence-bond Monte Carlo (VBMC) which can be used to efficiently sample ground states of quantum spin systems directly from the valence-bond basis.

In this Proceedings we describe some details of the methods used in our recent work on VBMC simulations of both uniform and random spin-1/2 antiferromagnetic Heisenberg chains and the closely related 1+1 dimensional quantum $Q$-state Potts models \cite{tra09}.  One motivation for this work is that for certain values of $Q$ these models describe interacting chains of non-Abelian anyons \cite{fei07}, exotic quasiparticle excitations which are thought to arise in certain fractional quantum Hall states \cite{nay08}. The results presented here are all for the case of uniform models (i.e., with no disorder) for which a number of exact results are known which can be used to benchmark the method.

\section{Hilbert space and models}

\begin{figure}[t]
  \begin{center}
  \includegraphics[width=6.5cm]{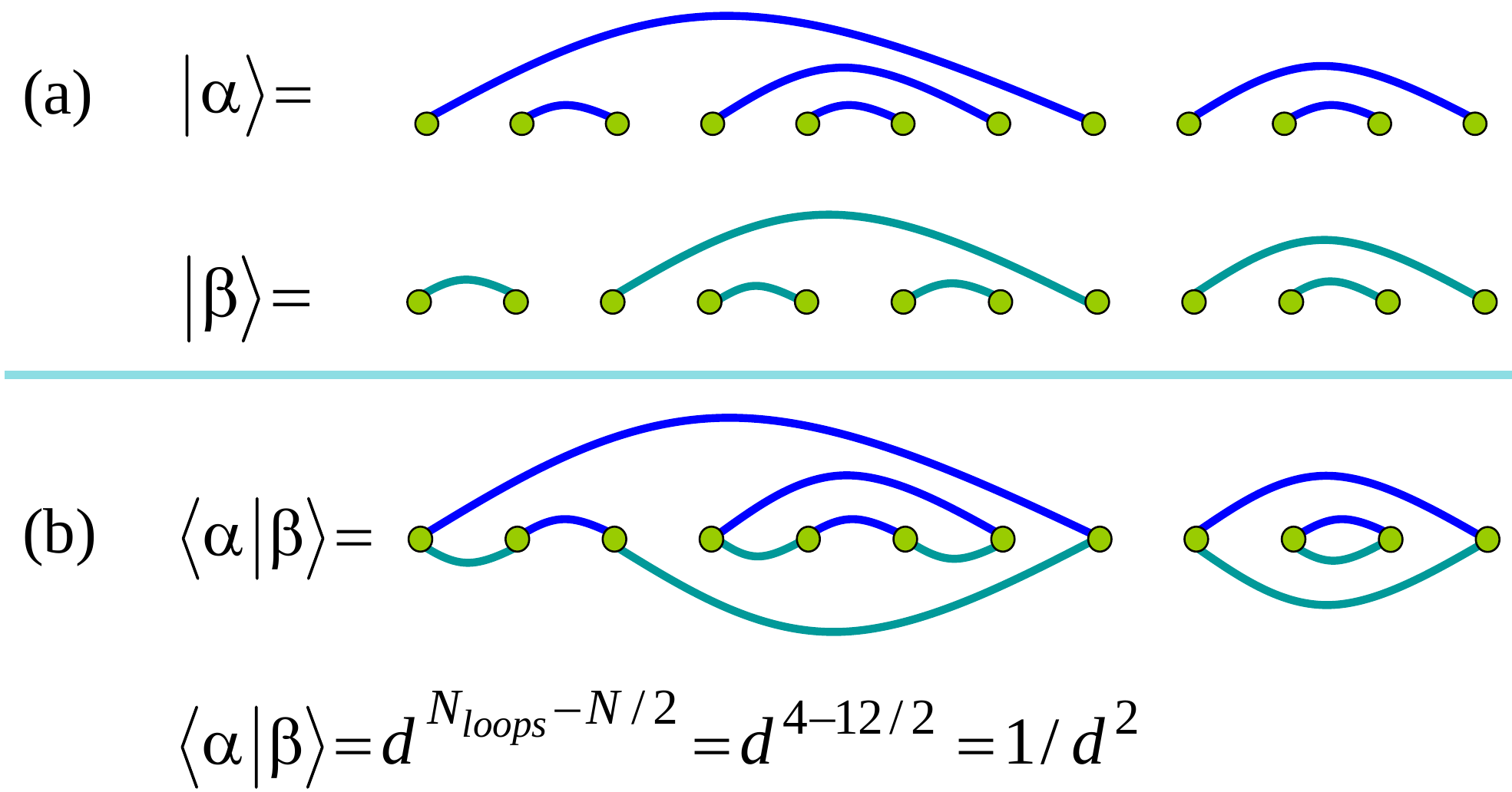}
  \caption{(a) Two normalized non-crossing valence-bond states $\vert \alpha \rangle$ and $\vert \beta \rangle$ in which pairs of particles (solid circles) are connected by valence bonds (solid lines). (b) To determine the overlap of these states one overlays the two valence-bond configurations and counts the number of closed loops, $N_{loops}$.  The overlap is then $\langle \alpha | \beta \rangle = d^{N_{loops} - N/2}$ where $d=2$ for the case of spin-1/2 particles.}\label{fig1}
  \end{center}
\end{figure}

We begin by describing the valence-bond basis for a chain of $N$ spin-1/2 particles.  This basis is made up of valence-bond states --- states in which all $N$ particles are paired up to form $N/2$ valence bonds.  Figure \ref{fig1}(a) shows two (normalized) valence-bond states $|\alpha\rangle$ and $|\beta\rangle$.  Both these states are examples of {\it non-crossing} valence-bond states, meaning that no two valence-bonds cross each other, or, equivalently, that the total spin of all the particles between any two particles connected by a valence bond must be 0.

For ordinary spin-1/2 particles the set of non-crossing valence-bond states forms a complete and {\it linearly independent} basis spanning the space of all total spin 0 states \cite{rum32}.  The number of these non-crossing states, and hence the dimensionality of the total spin 0 Hilbert space, grows asymptotically as $2^N$ for large $N$, as one would naturally expect for $N$ spin-1/2 particles.

One price to be paid for doing numerical calculations with the valence-bond basis is that it is a nonorthogonal basis.  The rule for determining the overlap of any two valence-bond states is shown in Fig.~\ref{fig1}(b) --- one simply overlays the two valence-bond configurations and counts the number of loops formed, $N_{loops}$.  The overlap is then $\langle \alpha| \beta \rangle = d^{N_{loops}-N/2}$, where, for spin-1/2 particles, $d=2$.

When $d\ne 2$ the Hilbert space spanned by non-crossing valence-bond states is still perfectly well defined provided $d \ge 1$, although it no longer describes a system of ordinary spin-1/2 particles.  For certain values of $d$, specifically when $d = 2 \cos \frac{\pi}{k+2}$ where $k$ is a positive integer, this Hilbert space can be interpreted physically as describing the `topological charge 0' sector of a system of $N$ non-Abelian anyons described by $su(2)_k$ Chern-Simons-Witten theory \cite{nay08}.  For these special values of $d$, the non-crossing valence-bond states are no longer linearly independent and the dimensionality of the Hilbert space for $N$ particles can be shown to grow asymptotically not as $2^N$ but as $d^N$.  The quantity $d$ is known as the quantum dimension of the particles \cite{kit06}.

Such $su(2)_k$ anyons are thought to arise physically in certain experimentally observed fractional quantum Hall states, notably the state with Landau-level filling fraction $\nu=5/2$ (corresponding to $k=2$) and, possibly, the $\nu=12/5$ state (corresponding to $k=3$) \cite{nay08}.

\begin{figure}[t]
  \begin{center}
  \includegraphics[width=6.5cm]{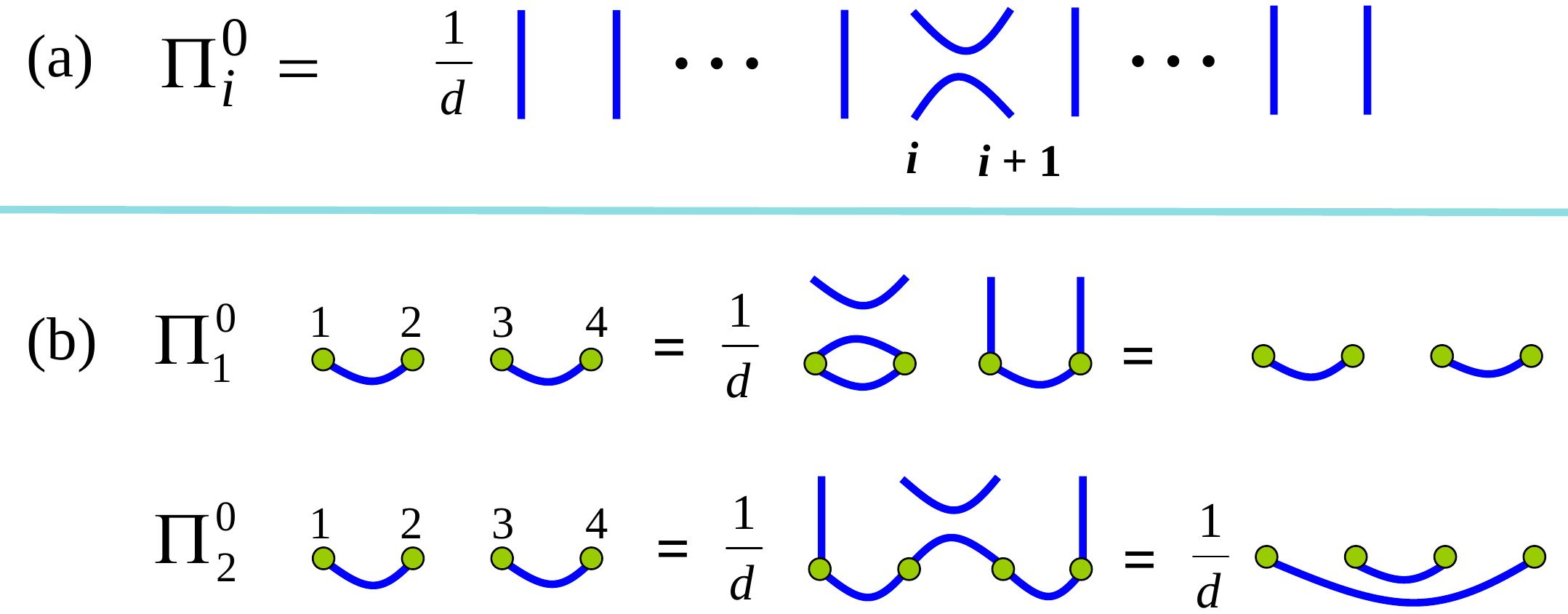}
    \caption{(a) Diagrammatic representation of a singlet projection operator, and (b) action of two projection operators on a given valence-bond state. In the first case (top), the loop formed introduces a factor of $d$ leading to an overall factor of $\frac{1}{d}\times d = 1$. In the second case (bottom), no loop is formed and the overall factor is $\frac{1}{d}$.} \label{fig2}
  \end{center}
\end{figure}

Having defined the relevant Hilbert spaces we now turn to the model Hamiltonians studied here.  To define these models we first describe the action of a nearest-neighbor singlet projection operator in the valence-bond basis.  Let $\Pi_i^0$ denote the singlet projection operator acting on sites $i$ and $i+1$.  Figure \ref{fig2}(a) shows a useful diagrammatic representation of this operator. Figure \ref{fig2}(b) uses this representation to illustrate the two distinct cases which can occur when acting on a valence-bond state with $\Pi_i^0$.  Either one applies the projection operator to two sites which are connected by a valence bond, in which case the projection operator has no effect on the state, or one acts on two sites which are each connected to different sites by valence bonds, in which case the projection operator forms a singlet between the two particles it acts on, as well as the two particles connected to them, and gives an overall factor of $1/d$.

The Hamiltonians we consider here all have the form
\begin{eqnarray}
H = - \sum_i \Pi_i^0. \label{model}
\end{eqnarray}
For ordinary spin-1/2 particles with $d=2$, the singlet projection operator can be expressed as $\Pi_i^0 = \frac{1}{4} - {\bf S}_i \cdot {\bf S}_{i+1}$ and (\ref{model}) corresponds to an antiferromagnetic nearest-neighbor Heisenberg chain. More generally, for arbitrary $d$, one can readily check that the operators $U_i = d~\Pi_i^0$ satisfy the so-called Temperley-Lieb algebra \cite{tem71},
\begin{eqnarray}
U_i^2 &=& d U_i,\cr
U_i U_{i\pm 1} U_i &=& U_i,\cr
[U_i,U_j] &=& 0,\ \ |i - j| > 1.
\end{eqnarray}
This algebra appears in the study of the 2-dimensional $Q$-state Potts model with $Q = d^2$, and as a consequence it can be shown that the models (\ref{model}) are equivalent to the 1+1 dimensional {\it quantum} $Q$-state Potts models  \cite{martinbook}.  Furthermore, for the special values $d= 2\cos\frac{\pi}{k+2}$ these models correspond to a sequence of conformally invariant Andrews-Baxter-Forrester \cite{and84} models with central charges $c_k = 1-6/(k+1)(k+2)$ \cite{hus84}.  As stated above, these models can be thought of as describing chains of interacting non-Abelian anyons \cite{fei07}.

\begin{figure*}[t]
  \begin{center}
  \includegraphics[width=6.5cm]{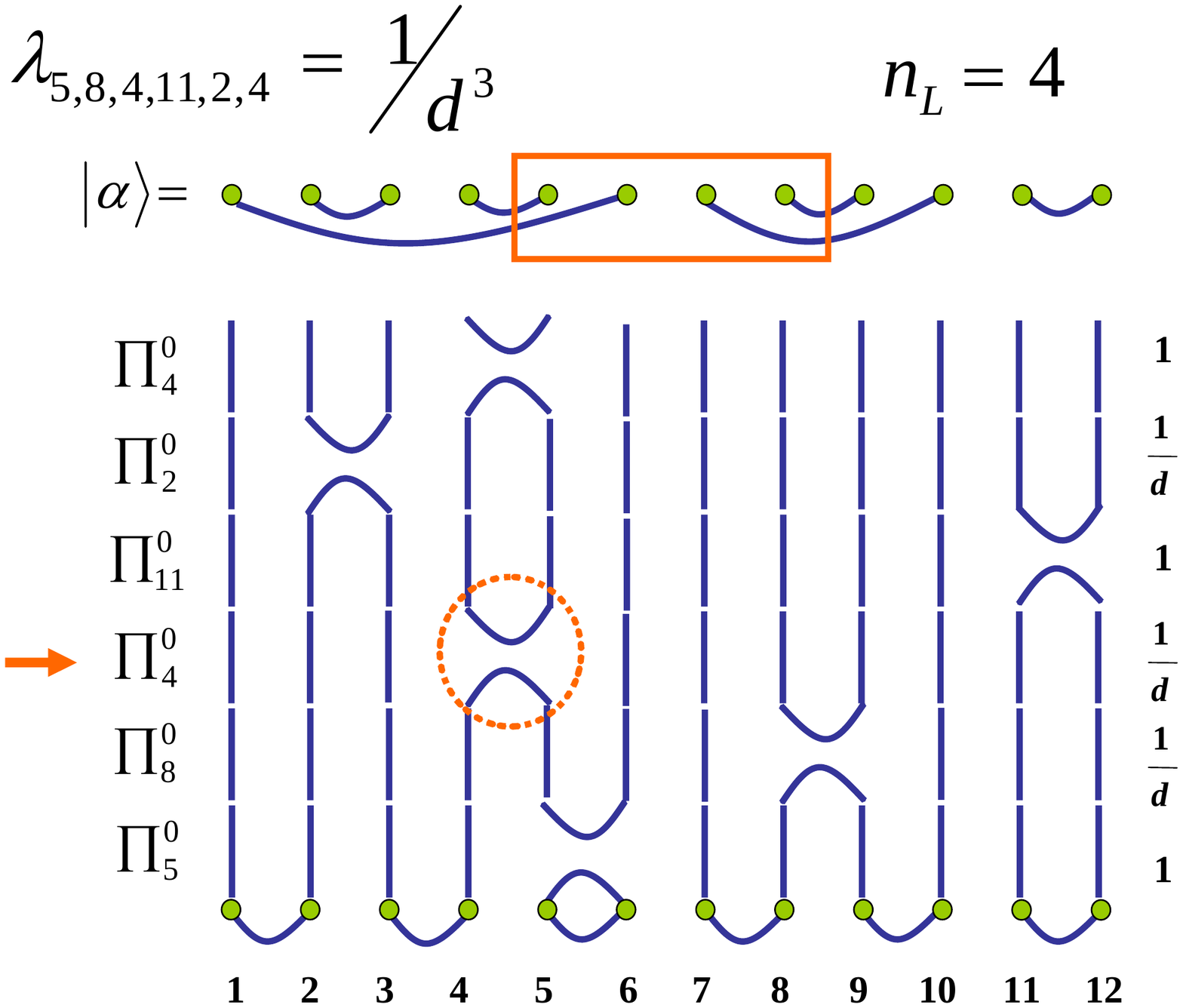}
  \includegraphics[width=6.5cm]{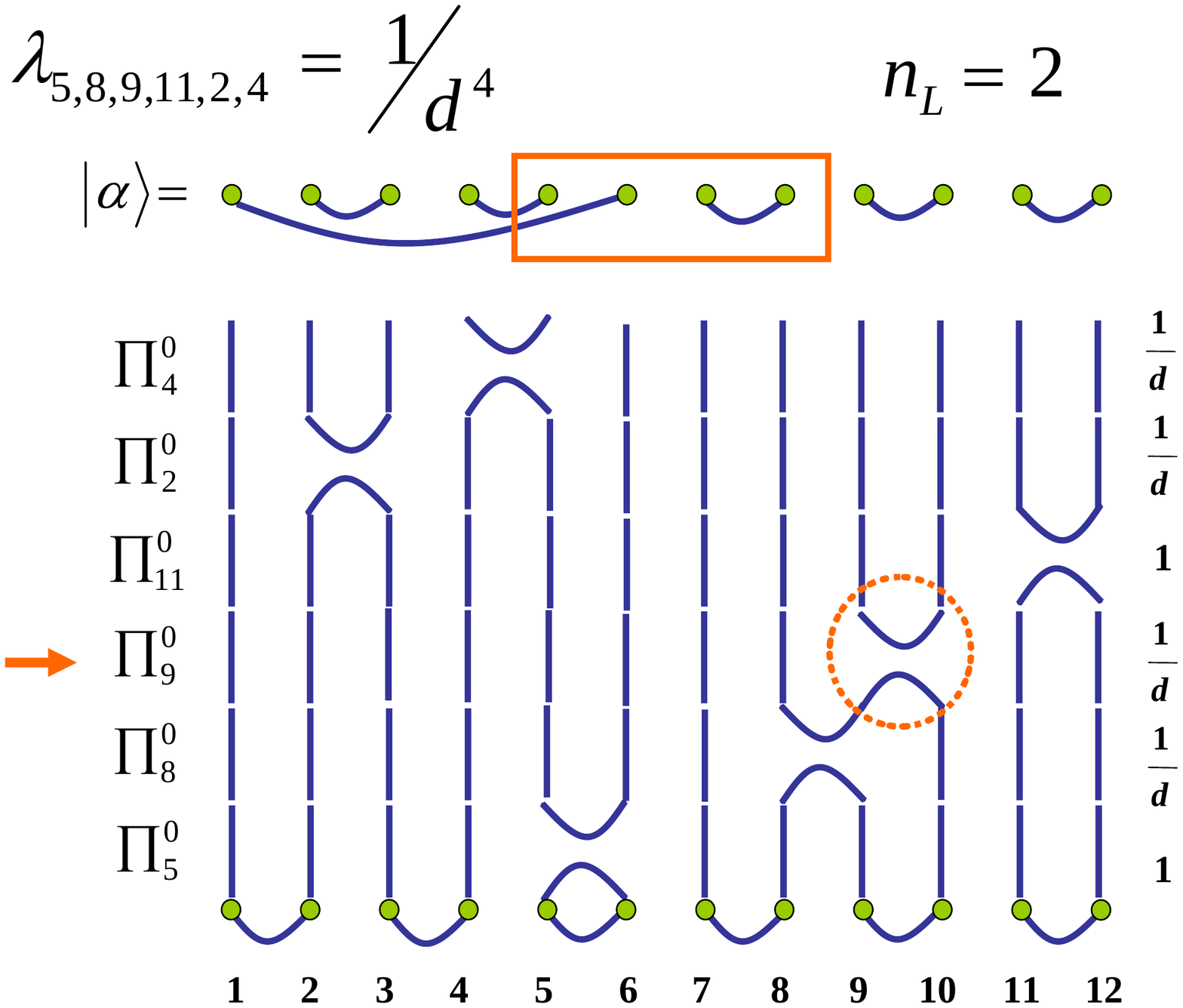}
   \caption{Diagrammatic representations of the action of two sequences of singlet projection operators on a given starting valence-bond state. In each diagram the resulting valence-bond state $\vert \alpha \rangle$ is determined by the open loops which terminate at the top of the diagram and the amplitude $\lambda_{i_1,\cdots,i_6}$ is the product of the factors listed on the right of the diagram. Each of these factors is either 1 or $1/d$, depending on whether the corresponding projection operator, listed on the left of the diagram, forms a closed loop or not.  The two diagrams shown can be viewed as `before' and `after' pictures for a single Monte Carlo update in VBMC in which the third projection operator from the bottom is shifted from $\Pi_4^0$ to $\Pi_9^0$ and the update is accepted or rejected according to the usual Metropolis rule. The quantity $n_L$ which is used to compute the valence-bond entanglement entropy (see text) is defined to be the total number of bond leaving a given block of $L$ sites (red rectangle).} \label{fig3}
  \end{center}
\end{figure*}

\section{Valence-bond Monte Carlo}

The basic idea behind VBMC is to act on a particular valence-bond state $\vert S\rangle$ repeatedly with $-H$ in order to project out the ground state. The results of this projection after $n$ iterations can be expressed as follows,
\begin{equation}
(-H)^n \vert S\rangle = \sum_{i_1,\cdots,i_n}  \Pi_{i_1}^0 \cdots \Pi_{i_n}^0 \vert S \rangle.\label{proj}
\end{equation}

Since acting on a non-crossing valence-bond state yields another non-crossing valence-bond state, (see Fig. \ref{fig2}(b)), it follows that
\begin{equation}
\Pi_{i_1}^0 \cdots \Pi_{i_n}^0 \vert S \rangle = \lambda_{i_1,\cdots,i_n} \vert \alpha\rangle \label{term}
\end{equation}
where $\vert \alpha\rangle$ is a non-crossing valence-bond state with the same norm as $\vert S \rangle$ and $\lambda_{i_1, \cdots, i_n} = d^{-m}$ where $m$ is the number of time a projection operator acts on two sites which are not connected by a valence bond in the process of projecting $|S\rangle$ onto the state $|\alpha\rangle$.   Figure \ref{fig3} shows diagrammatic representations for two different terms of the form (\ref{term}) with $N=12$ and $n=6$ (in our simulations we study system sizes up to $N = 1024$ and take $n = 20 N$).

In the end the projection (\ref{proj}) leads to an expression for the ground state $|\psi\rangle$ of the form
\begin{eqnarray}
|\psi\rangle = \sum_\alpha w(\alpha) |\alpha\rangle.\label{wf}
\end{eqnarray}
In VBMC one samples the valence-bond states $|\alpha\rangle$ contributing to $|\psi\rangle$ with probability $w(\alpha)/\sum_\beta w(\beta)$ by updating the sequence of projection operators $\Pi_{i_1}^0,\cdots,\Pi_{i_n}^0$ using the usual Metropolis method.  One such Monte Carlo update is shown in Fig.~\ref{fig3}.

Given any observable $O$ with expectation values $O(\alpha) = \langle \alpha |O| \alpha \rangle$ in the non-crossing valence-bond states $|\alpha\rangle$, VBMC can be used to compute the average $\langle O \rangle = \sum_\alpha w(\alpha) O(\alpha)/\sum_\alpha w(\alpha)$ for any state $|\psi\rangle$ of the form (\ref{wf}), provided $w(\alpha) \ge 0$ (which is the case here).  In what follows, angle brackets will always denote this average, though it should be noted that $\langle O \rangle$ will in general not be equal to the true expectation value $\langle \psi|  O |\psi\rangle/\langle \psi| \psi \rangle$, both because the valence-bond states are nonorthogonal and because the weight factors $w(\alpha)$ are amplitudes and not probabilities.  Of course the true quantum expectation value of any operator can be computed by VBMC if one carries out the projection on both the bra state and the ket state \cite{san05}.  However, here we focus on those quantities which can be calculated efficiently using the ``one-sided" VBMC described above.

\section{Results}

\subsection{Ground state energy}

One quantity which is easily computed using one-sided VBMC is the ground state energy $E_0$.  The procedure given in \cite{san05} for calculating $E_0$ for spin-1/2 systems can be trivially generalized for arbitrary $d$ and leads to the following expression,
\begin{equation}
\frac{E_0}{N}=-\frac{1}{N}\left\langle \sum_{i=1}^N w_i\right\rangle.\label{energy}
\end{equation}
Here, $w_i$ is equal to 1 if sites $i$ and $i+1$ are connected by a valence bond and $1/d$ if they are not for a given valence-bond state $|\alpha\rangle$, and, as described above, the angle brackets denote an average over these valence-bond states weighted by the amplitudes $w(\alpha)$.

In fact, the ground state energies of the models (\ref{model}) can be found exactly.  This can be seen by noting that for any $d$ the Temperley-Lieb operators can be represented using spin-1/2 operators as $U_i = 2(S_i^x S_{i+1}^x + S_i^y S_{i+1}^y) + d (1/4 - S_i^z S_{i+1}^z) + i\sqrt{1-d^2/4} (S^z_{i+1} - S^z_i)$ \cite{martinbook}.  For the case of open boundary conditions the models (\ref{model}) can then be mapped onto spin-1/2 XXZ chains with external (non-Hermitian) fields applied to the two ends (the staggered field term in the expression for $U_i$ cancels in the ``bulk" of the chain).  In the thermodynamic limit, the ground state energies will not depend on boundary conditions, and the values of $E_0$ for the models (\ref{model}) with periodic boundary conditions should be the same as that for the corresponding $XXZ$ models.  Using the expression for the ground state energies of the $XXZ$ models found using Bethe ansatz by Yang and Yang \cite{yan66} it is straightforward to obtain the following expression for the ground energies of the models (\ref{model}),
\begin{equation}
\frac{E_0}{N}=\frac{d^2-4}{4d}\int_{-\infty}^\infty
dx\frac{{\rm sech}(\pi
x)}{\cosh\left(2x\arccos\frac{d}{2}\right)-\frac{d}{2}}.\label{exact}
\end{equation}

\begin{figure}[t]
  \begin{center}
  \includegraphics[width=6.5cm]{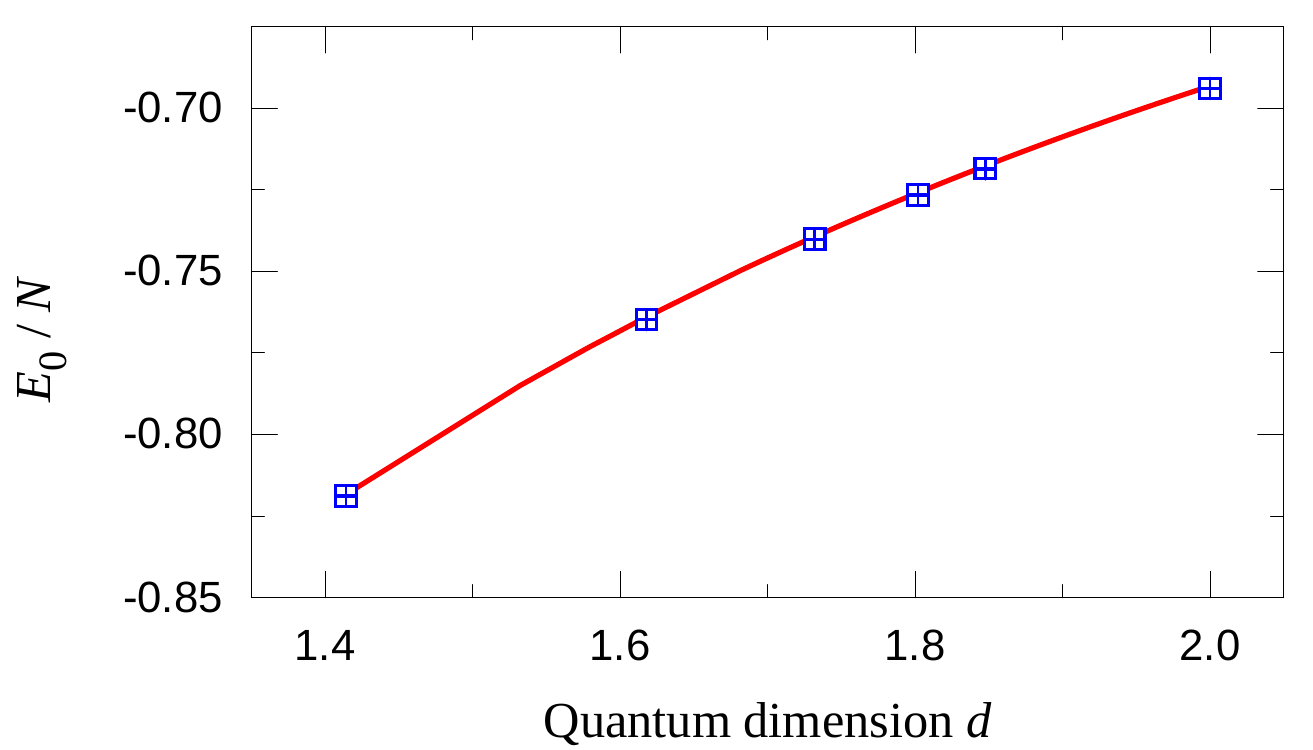}
   \caption{Ground state energy per site $E_0/N$ as a function of quantum dimension $d$.  The solid red line is the exact Bethe ansatz result, and the blue squares are the results of our VBMC simulations (error bars are smaller than symbol size).} \label{fig4}
  \end{center}
\end{figure}

Figure \ref{fig4} shows the ground state energies we obtained by evaluating the expression (\ref{energy}) by VBMC for $d = 2 \cos \frac{\pi}{k+2}$ with $k=2,3,4,5,6$ and $\infty$.  The red line is the exact energy (\ref{exact}) as a function of $d$. The fact that our numerical results clearly agree with the exact Bethe ansatz results should be seen as evidence that VBMC can indeed be used to simulate the models (\ref{model}) with arbitrary $d$.

\begin{figure*}[t]
  \begin{center}
    \includegraphics[width=13cm]{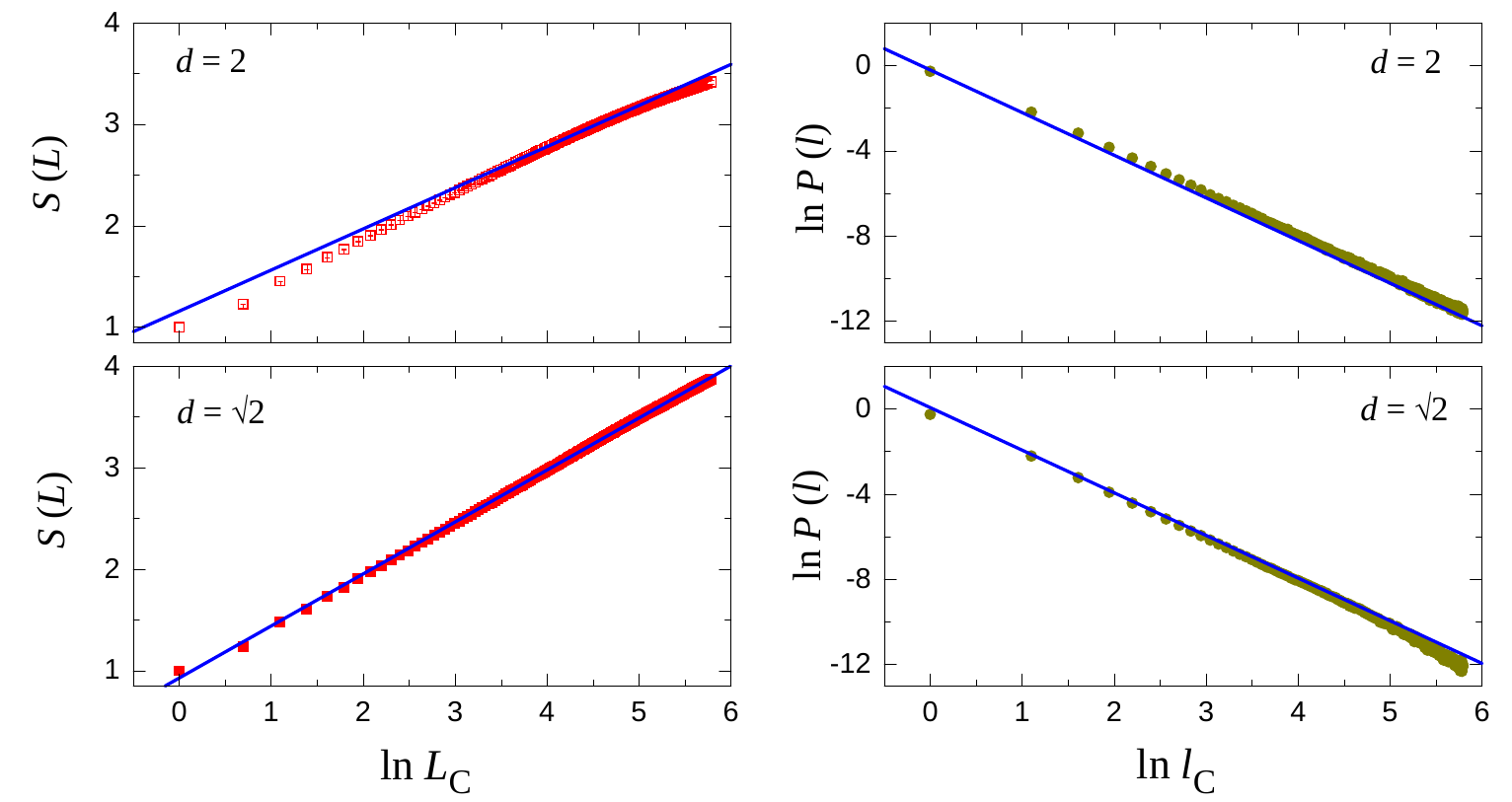}
  \caption{Semi-log plots of the valence-bond entanglement entropy $S(L)$ as functions of the block size $L_c$, and log-log plots of the bond-length distribution $P(l)$ as a function of bond length $l_c$ for $d=\sqrt{2}$ and $d=2$. Here $L_c = (N/\pi) \sin (\pi L/N)$ and $l_c = (N/\pi) \sin(\pi l/N)$ are the so-called conformal lengths which we use to minimize finite size effects when $L$ and $l$ are nearly equal to half the system size.  The solid lines in these plots are the analytic results for the asymptotic scaling of $S(L)$ and $P(l)$ which follow from the results  obtained in \cite{jac08}.  Results are for chains with $N=1024$ sites and the parameter $n$ is taken to be $20 N$.}
   \label{fig5}
   \end{center}
\end{figure*}

\subsection{Bond length distribution and valence-bond entanglement}

Another quantity which is natural to compute using one-sided VBMC is the so-called valence-bond entanglement entropy \cite{ale07,chh07}.  If $n_L$ is defined to be the total number of bonds leaving a contiguous block of $L$ sites in a given valence-bond state $|\alpha\rangle$ (see Fig.~\ref{fig3}) then the valence-bond entanglement entropy is defined to be $S(L) = \langle n_L \rangle$.  While this quantity was originally only defined for spin-1/2 systems, the definition clearly generalizes to the models considered here for arbitrary values of $d$, as first noted by Jacobsen and Saleur \cite{jac08}.  In this same paper, Jacobsen and Saleur also obtained analytic results for the $L \gg 1$ scaling of $S(L)$.  For all $d \le 2$ they found that $S(L)$ scales logarithmically with $L$ with a $d$ dependent coefficient.  It should be noted that this exact result was based on a mapping to a long-wavelength field theoretic description of the models, and so it is worthwhile testing this prediction numerically.

Before describing our results for $S(L)$, we note that this quantity is closely related to the bond-length distribution, $P(l)$, for valence bonds. This distribution is defined to be $P(l) =  \langle b_l \rangle/(N/2)$ where $b_l$ is the number of valence bonds of length $l$ in a given valence-bond state $\vert \alpha \rangle$.   It is readily shown \cite{hoyos07} that $P(l)$ is related to $S(L)$ by,
\begin{equation}
\label{nLpl}
S(L) = 2 \sum_{l=1}^{N} P(l) \min(l,L).
\end{equation}
(Note that $P(l) = 0$ for even $l$.) This expression, together with the result of \cite{jac08} that $S(L)$ scales logarithmically with $L$ for $d \le 2$ implies that, for these values of $d$, $P(l)$ should follow an inverse-square power law for $l \gg 1$.

Figure \ref{fig5} shows our VBMC results for $S(L)$ and $P(l)$ for the cases $d=\sqrt 2$ (corresponding to the critical one-dimensional transverse field Ising model) and $d=2$ (corresponding to the spin-1/2 antiferromagnetic Heisenberg chain).  The log-log plots of $P(l)$ vs. $l$ clearly demonstrate the predicted inverse-square power law dependence for the bond-length distribution.  The semi-log plots of $S(L)$ also show the expected logarithmic scaling of the valence-bond entanglement entropy, for $L \gg 1$.

\section{Conclusions}

To summarize, we have shown that the VBMC method of Sandvik \cite{san05} can be straightforwardly generalized to study the 1+1 dimensional quantum $Q(=d^2)$-state Potts models.  For $d=2 \cos\frac{\pi}{k+2}$ these models describe chains of interacting non-Abelian anyons, exotic quasiparticle excitations believed to exist in certain experimentally observed fractional quantum Hall states \cite{fei07,nay08}.  The ground state energies, bond-length distributions and valence-bond entanglement entropies of these models were computed using VBMC and compared to various known exact results.  This work sets the stage for our recent VBMC study of the effect of disorder on these models \cite{tra09}.

\section{Acknowledgments}
We thank S. H. Simon for useful discussions.  This work is supported by US DOE Grant No. DE-FG02-97ER45639. Computational work was performed at the Florida State University High Performance Computing Center.


\begin{thebibliography}{}
\bibitem{lia88} S. Liang, B. Dou{\c c}ot, and P. W. Anderson, Phys. Rev. Lett. \textbf{61}, 365 (1988).

\bibitem{san05} A. W. Sandvik, Phys. Rev. Lett. \textbf{95}, 207203 (2005).

\bibitem{tra09} H. Tran and N. E. Bonesteel, Preprint, arXiv: 0909.0038.

\bibitem{fei07} A. Feiguin \emph{et al.}, Phys. Rev. Lett. \textbf{98} 160409 (2007).

\bibitem{nay08} C. Nayak \emph{et al.}, Rev. Mod. Phys. \textbf{80}, 1083 (2008).

\bibitem{rum32} G. Rumer, G\"ottingen Nachr. Tech. {\bf 1932}, 377 (1932).

\bibitem{kit06} A. Yu. Kitaev, Ann. Phys. (N.Y.) {\bf 321}, 2 (2006).

\bibitem{tem71} H. N. V. Temperley and E. H. Lieb, Proc. R. Soc. London, Ser. A {\bf 322}, 251 (1971).

\bibitem{martinbook} P. P. Martin, {\it Potts Models and Related Problems in Statistical Mechanics} (World Scientific, 1991).

\bibitem{and84} G. E. Andrews, R. J. Baxter and P. J. Forrester, J. Stat. Phys. \textbf{35}, 193 (1984).

\bibitem{hus84} D. A. Huse, Phys. Rev. B \textbf{30}, 3908 (1984)

\bibitem{yan66} C. N. Yang and C. P. Yang, Phys. Rev. {\bf 150}, 321 (1966); Phys. Rev. {\bf 150}, 327 (1966).

\bibitem{ale07} F. Alet \emph{et al.}, Phys. Rev. Lett. \textbf{99}, 117204 (2007).

\bibitem{chh07} R. W. Chhajlany, P. Tomczak, and A. W\'{o}jcik, Phys. Rev. Lett. \textbf{99}, 167204 (2007).

\bibitem{jac08} J. L. Jacobsen and H. Saleur, Phys. Rev. Lett. {\bf 100}, 087205 (2008).

\bibitem{hoyos07} J. A. Hoyos \emph{et al.}, Phys. Rev. B \textbf{76}, 174425 (2007).

\end{thebibliography}
\end{document}